# Comment on "Higher order reentrant post modes in cylindrical cavities" [J. App. Phys. 122, 144501 (2017)]

S. Belomestnykh
*Fermi National Accelerator Laboratory, Batavia, IL 60510-5011, USA*

*Abstract*

A recent article [McAlister *et al.*, J. Appl. Phys. **122**, 144501 (2017)] claims to show the existence of "new" higher order reentrant post modes. Indeed, such modes do exist and are known to scientists and engineers for a very long time as modes of a foreshortened quarter-wave coaxial resonator. It is a textbook problem. In this comment, I briefly review the higher order modes in such cavities and provide relevant references.

McAllister *et al.*[1] recently published a study of what they consider "new" higher order reentrant post modes in cylindrical cavities. In fact, the authors have studied a textbook problem of a coaxial quarter-wave resonator and its transition to a cylindrical cavity. Coaxial and reentrant cavities are widely used in particle accelerators[2] to provide acceleration or transverse kicking of charged particles[3,4]; their higher order modes have been studied extensively; and these cavities are taught about in various courses, e.g. Ref. 5. The cavities of this type, very often filled with dielectric materials to reduce their dimensions, are also used as filters if radio-frequency circuits. There are well-developed techniques for their analysis, design and measurements.

With a zero gap, such cavity is simply a half-wave resonator. That is, transverse electromagnetic (TEM) resonant standing-wave modes occur at wavelengths, when the enclosed length $l$ is an integer multiple of half wavelengths. (In this comment, I do not consider higher order modes other than monopole.) both the electric and magnetic fields have only transverse components, $E_r$ and $B_\phi$ correspondingly[3]:

$$E_r = E_0 \sin(p\pi z/l)\, e^{-i\omega t}. \tag{1}$$

$$B_\phi = B_0 \cos(p\pi z/l)\, e^{-i\omega t}, \tag{2}$$

These formulas describe fields with *FF* = 0 presented in Figures 4 and 5, Ref. 1. The fields are of pure TEM nature. While the authors of Ref. 1 recognize the modes as coaxial, they failed to identify them as TEM and make connection to the field in cavities with a non-zero gap.

As the gap opens, the cavity becomes a foreshortened quarter-wave resonator or QWR (when the gap is small with respect to the wavelength), and then slowly transitions to a reentrant and eventually to a simple cylindrical cavity. This is evident from comparing Figure 1 below with Figure 6(a) in Ref. 1. As the gap opens up, the mode frequencies of the foreshortened quarter-wave resonator increase. When one of the QWR frequencies nears the frequency of one of the cylindrical cavity $TM_{01n}$ modes, the quarter-wave cavity mode begins to transition into the latter mode. One might call this "transition" mode the reentrant cavity mode. When the lower frequency QWR mode approaches the same $TM_{01n}$ mode, it pushes the higher order reentrant cavity mode up to the higher order $TM_{01n}$ mode. So eventually all cylindrical cavity modes are occupied.

The foreshortened quarter-wave resonator can be modeled as a coaxial line shorted at one end terminated by a capacitance (a gap between the center conductor and the conducting end wall) at the other end. The resonant condition can be written as

$$\omega C_0 Z_0 - \cot\left(\frac{2\pi l}{\lambda}\right) = 0, \tag{3}$$

where $C_0$ is the gap capacitance, $Z_0$ is the characteristic impedance of the transmission line, $\lambda$ is the wavelength, and $\omega$ is the frequency. The solution of this equation gives us resonant wavelengths of the cavity modes. When the gap capacitance is small, the modes have the wavelengths of approximately

$$\lambda_n \approx 4l/(2n+1), \tag{4}$$

where n = 0, 1, 2, … so that the cavity length is equal to $\lambda_0/4$, $3\lambda_1/4$, $5\lambda_2/4$, etc. The quarter-wave resonator modes have field pattern of a standing TEM wave in the regular part of coaxial line and capacitive character near the gap. These are the modes that are referred to as "reentrant post modes" or "new modes" in Ref. 1. The authors claim[1] "… we have demonstrated existence of higher order reentrant modes with higher frequencies than the fundamental …" First, all higher order modes have higher frequencies than the fundamental mode by definition. Second, there is no need to demonstrate "existence" of these modes as they have been known for a very long time.

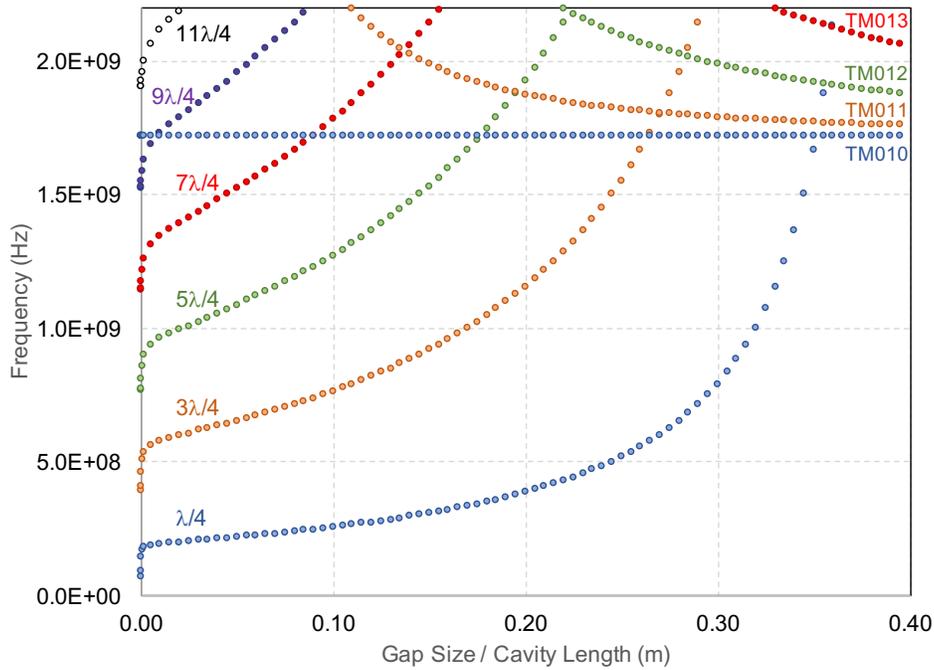

Figure 1: Frequencies of foreshortened quarter-wave resonator modes and $TM_{01n}$ modes of cylindrical cavity as a function of gap size and cavity length respectively. Here a very simple model of the gap capacitance is used for the quarter-wave resonator calculations. The cylindrical cavity modes are assumed to occupy only the gap area.

The authors' statement that "the geometry factor … is directly proportional to the modes' quality factor", while formally correct, is misleading. The geometry factor is determined by the cavity geometry only and does not depend on the cavity frequency or material properties. So, for a given geometry it is a constant. On the other hand, the quality factor does depend strongly on the frequency and material properties via the surface resistivity.

In addition, the experimental setup described in Ref. 1 lacked proper RF contact (e.g. a sliding spring contact or a choke joint) between the post and the cavity lid. As a result, the measured quality factor values have extremely poor agreement with the calculated values.

In conclusion, if the authors limited themselves to the subject of studying behavior of the form factor *C* on the cavity geometry and finding an optimal cavity configuration(s) for axion detection, it would be OK. However, they made a point in the title and abstract that the results are of a more fundamental nature. The authors' claim that they have demonstrated the existence of "new" modes is incorrect as such modes has been known and studied for a very long time. The measurement results presented in the article [1] are of a poor quality.

*Acknowledgement*